\documentclass[10pt]{article}

\usepackage{amsmath} 
\usepackage{graphicx} 
\usepackage{labelfig}

\begin{document}

\baselineskip=4.6mm

\makeatletter

\newcommand{\E}{\mathrm{e}\kern0.2pt}
\newcommand{\D}{\mathrm{d}\kern0.2pt}
\newcommand{\RR}{\mathrm{I\kern-0.20emR}}

\def\bottomfraction{0.9}

\title{{\bf Bounds for solutions to the problem \\ of steady water waves with
vorticity}}

\author{Vladimir Kozlov$^a$, Nikolay Kuznetsov$^b$}

\date{}

\maketitle

\vspace{-10mm}

\begin{center}
$^a${\it Department of Mathematics, Link\"oping University, S--581 83 Link\"oping,
Sweden \\ $^b$ Laboratory for Mathematical Modelling of Wave Phenomena, \\ Institute
for Problems in Mechanical Engineering, Russian Academy of Sciences, \\ V.O.,
Bol'shoy pr. 61, St. Petersburg 199178, Russian Federation} \\

\vspace{2mm}

E-mail: vladimir.kozlov@liu.se\,/\,nikolay.g.kuznetsov@gmail.com
\end{center}

\begin{abstract}
The two-dimensional free-boundary problem describing steady gravity waves with
vorticity on water of finite depth is considered. Bounds for stream functions as
well as free-surface profiles and the total head are obtained under the assumption
that the vorticity distribution is a locally Lipschitz function. It is also shown
that wave flows have counter-currents in the case when the infimum of the free
surface profile exceeds a certain critical value.
\end{abstract}

\section{Introduction}

\setcounter{equation}{0}

We consider the two-dimensional nonlinear problem describing steady waves in a
horizontal open channel occupied by an inviscid, incompressible, heavy fluid, say
water. The water motion is assumed to be rotational which, according to
observations, is the type of motion commonly occurring in nature (see, for example,
\cite{SCJ,Th} and references cited therein). A brief characterization of various
results concerning waves with vorticity on water of finite as well as infinite depth
is given in \cite{KK2}. Further details can be found in the survey article
\cite{WS}.

Studies of bounds on characteristics of waves with vorticity were initiated by Keady
and Norbury almost 40 years ago in the pioneering paper \cite{KN}. In our
article~\cite{KK4}, we continued these studies and extended the results of \cite{KN}
to all types of vorticity distributions. In the recent note \cite{KKL}, it was
demonstrated that restrictions on the Lipschitz constant of the vorticity
distribution imposed in \cite{KN} and \cite{KK4} are superfluous in the case of
unidirectional flows. Our aim here is to obtain the same bounds as in \cite{KK4}
under the assumption that the vorticity distribution is just locally Lipschitz
continuous. Moreover, we show that wave flows have counter-currents in the case when
the infimum of the free surface profile exceeds a certain critical value; the latter
is equal to the largest depth that have unidirectional, shear flows with horizontal
free surfaces (see formula \eqref{eq:h_0} below).

The plan of the paper is as follows. Statement of the problem is given in \S\,1.1
and some background facts are listed in \S\,1.2. Necessary facts about auxiliary
one-dimensional problems are presented in \S\,1.3 (see further details in
\cite{KK3}), whereas main results are formulated in \S\,1.4. Two auxiliary lemmas
are proved in \S\,2, whereas proofs of Theorems~1--4 are given in \S\,3. In the last
section, we discuss some improvement of Theorem~1 that follows when a rather weak
condition is imposed on the derivative of the vorticity distribution instead of the
restriction required in Theorem~3.2, \cite{KK4}.

\subsection{Statement of the problem}

Let an open channel of uniform rectangular cross-section be bounded below by a
horizontal rigid bottom and let water occupying the channel be bounded above by a
free surface not touching the bottom. The surface tension is neglected and the
pressure is constant on the free surface. The water motion is supposed to be
two-dimensional and rotational which combined with the water incompressibility
allows us to seek the velocity field in the form $(\psi_y, -\psi_x)$, where $\psi
(x,y)$ is referred to as the {\it stream function} (see, for example, the book
\cite{LSh}). The vorticity distribution $\omega$ is prescribed and depends on $\psi$
as is explained in \cite{LSh}, \S~1. The vorticity distribution is supposed to be
locally Lipschitz continuous, but we impose no condition on the Lipschitz constant
here which distinguishes the present article from \cite{KN} and \cite{KK4}.
Moreover, unlike the recent note \cite{KKL} wave flows with counter-currents are
considered here along with unidirectional ones.

Furthermore, our results essentially use the following classification of vorticity
distributions which is based on properties of $\Omega (\tau) = \int_0^\tau \omega
(t) \, \D t$ and slightly differs from that proposed in \cite{KK3}:

\vspace{1mm}

\noindent \ \ (i) $\max_{0 \leq \tau \leq 1} \Omega (\tau)$ is attained either at an
inner point of $(0, 1)$ or at one (or both) of the end-points. In the latter case,
either $\omega (1) = 0$ when $\Omega (1) > \Omega (\tau)$ for $\tau \in (0, 1)$ or
$\omega (0) = 0$ when $\Omega (0) > \Omega (\tau)$ for $\tau \in (0, 1)$ (or both of
these conditions hold simultaneously).

\noindent \ (ii) $\Omega (0) > \Omega (\tau)$ for $\tau \in (0, 1]$ and $\omega (0)
< 0$.

\noindent (iii) $\Omega (\tau) < \Omega (1)$ for $\tau \in (0, 1)$ and $\omega (1) >
0$. Moreover, if $\Omega (1) = 0$, then $\omega (0) < 0$ and $\omega (1) > 0$ must
hold simultaneously.

\vspace{1mm}

\noindent It should be noted that conditions (i)--(iii) define three disjoint sets
of vorticity distributions whose union is the set of all continuous distributions on
$[0,1]$.

Non-dimensional variables are used with lengths and velocities scaled to
$(Q^2/g)^{1/3}$ and $(Qg)^{1/3}$, respectively; here $Q$ and $g$ are the dimensional
quantities for the volume rate of flow per unit span and the constant gravity
acceleration respectively. (We recall that $(Q^2/g)^{1/3}$ is the depth of the
critical uniform stream in the irrotational case; see, for example, \cite{Ben}.)
Hence the constant rate of flow and the acceleration due to gravity are scaled to
unity in our equations.

In appropriate Cartesian coordinates $(x,y)$, the bottom coincides with the $x$-axis
and gravity acts in the negative $y$-direction. We choose the frame of reference so
that the velocity field is time-independent as well as the unknown free-surface
profile. The latter is assumed to be the graph of $y = \eta (x)$, $x\in \RR$, where
$\eta$ is a bounded positive $C^1$-function. Therefore, the longitudinal section of the
water domain is
\[ D = \{ x \in \RR,\ 0 < y < \eta (x) \} .
\]
Since the surface tension is neglected, the pair $(\psi , \eta)$ with $\psi \in C^2
(D) \cap C^{1} (\bar D)$ must satisfy the following free-boundary problem:
\begin{eqnarray}
&& \psi_{xx} + \psi_{yy} + \omega (\psi) = 0, \quad (x,y)\in D;
\label{eq:lapp} \\ && \psi (x,0) = 0, \quad x \in \RR; \label{eq:bcp} \\ && \psi
(x,\eta (x)) = 1, \quad x \in \RR; \label{eq:kcp} \\ && |\nabla \psi (x,\eta (x))|^2
+ 2 \eta (x) = 3 r, \quad x \in \RR . \label{eq:bep}
\end{eqnarray}
Here, the constant $r$ is problem's parameter referred to as the total head or the
Bernoulli constant (see, for example, \cite{KN}). Throughout the paper we assume
that
\begin{equation}
|\psi| \ \mbox{is bounded and} \ |\psi_x| , \ |\psi_y| \ \mbox{are bounded and
uniformly continuous on} \ \bar D . \label{bound}
\end{equation}
The formulated statement of the problem has long been known and its derivation from
the governing equations and the assumptions about the boundary behaviour of water
particles can be found, for example, in \cite{CS}. Notice that the boundary
condition \eqref{eq:kcp} yields that relation \eqref{eq:bep} (Bernoulli's equation)
can be written as follows:
\begin{equation*}
\left[ \partial_n \psi (x,\eta (x)) \right]^2  + 2 \eta (x) = 3 r, \quad x \in \RR
\, . \label{eq:ben}
\end{equation*}
By $\partial_n$ we denote the normal derivative on $\partial D$, where the unit
normal $n = (n_x, n_y)$ points out of $D$.

\subsection{Background}

We begin with results obtained in the irrotational case, an extensive description of
which one finds in \cite{Ben}, \S\,1. Therefore, we restrict ourselves only to the
most important papers. As early as 1954, Benjamin and Lighthill \cite{BenL}
conjectured that $r > 1$ for all steady wave trains on irrotational flows of finite
depth. For a long period only two special kinds of waves were known, namely, Stokes
waves (periodic waves whose profiles rise and fall exactly once per period), and
solitary waves (such a wave has a pulse-like profile that is symmetric about the
vertical line through the wave crest and monotonically decreases away from it). The
inequality $r > 1$ for Stokes waves was proved by Keady and Norbury \cite{KN1} (see
also Benjamin \cite{Ben}), whereas Amick and Toland \cite{AT} obtained the proof for
solitary waves. Finally, Kozlov and Kuznetsov \cite{KK0} proved this inequality
irrespective of the wave type (Stokes, solitary, whatever) under rather weak
assumptions about wave profiles; in particular, stagnation points are possible on
them.

Furthermore, estimates of 
\[ \hat{\eta} = \sup_{x \in \RR} \eta (x) \quad \mbox{and} \quad \check{\eta} =
\inf_{x \in \RR} \eta (x) 
\]
were found for Stokes waves in the paper \cite{KN1}. They are expressed in terms of
the depths of the supercritical and subcritical uniform streams. Benjamin had
recovered these estimates in his article \cite{Ben}, in which the inequality for
$\check{\eta}$ is generalised to periodic waves that bifurcate from the Stokes ones.
(It should be noted that only numerical evidence indicated their existence in 1995
when \cite{Ben} was published and, to the authors knowledge, there is no rigorous
proof up to the present.) For arbitrary steady wave profiles natural estimates of
these quantities were obtained in \cite{KK0} under the same assumptions as the
inequality $r > 1$. Namely, it was shown that $\check{\eta}$ is strictly less than
the depth of the subcritical uniform stream, whereas $\hat{\eta}$ is strictly
greater than it. Also, an arbitrary wave profile lies strictly above the horizontal
surface of the supercritical uniform stream, but it is well known that profiles of
solitary waves asymptote the latter at infinity.

Now we turn to the case of waves with vorticity considered in the framework of
problem \eqref{eq:lapp}--\eqref{eq:bep}. The first paper relevant to cite in this
connection was the paper \cite{KN} by Keady and Norbury who, in particular,
generalised their estimates of $\check{\eta}$ and $\hat{\eta}$ obtained for
irrotational waves in \cite{KN1}. It should be emphasised that for this purpose they
subject the vorticity distribution $\omega$ to rather strong assumptions, one of
which restricted their considerations only to distributions that satisfy conditions
(i) (see details in \S\,4). In our article \cite{KK4}, this restriction was
eliminated, whereas another one was still used. Unfortunately, some assumptions
required for proving a couple of assertions are missing in \cite{KK4} (see details
in \S\,4). On the other hand, various superfluous requirements imposed in that
paper, in particular, on the derivative of the vorticity distribution were
eliminated in the note \cite{KKL}, but only in the case of unidirectional flows.

\subsection{Auxiliary one-dimensional problems}

First, we outline some properties of solutions to the equation
\begin{equation}
U'' + \omega (U) = 0 , \ \ \ y \in \RR ;
\label{eq:U}
\end{equation}
here and below $'$ stands for $\D / \D y$. These results were obtained in \cite{KK3}
and are essential for our considerations.

The set of solutions is invariant with respect to the following transformations: $y
\mapsto y + \mbox{constant}$ and $y \mapsto - y$. There are three immediate
consequences of this property: (a) if a solution of equation \eqref{eq:U} has no
stationary points, then it is strictly monotonic (either increasing or decreasing)
on the whole $y$-axis; (b) if a solution has a single stationary point, say $y=y_0$,
then the solution's graph is symmetric about the straight line through $y=y_0$ that
is orthogonal to the $y$-axis, the solution decreases (increases) monotonically in
both directions of the $y$-axis away from this point provided it attains its maximum
(minimum respectively) there; (c) if a solution has two stationary points, then
there are infinitely many of them and the solution is periodic with one maximum and
one minimum per period, it is monotonic between the extrema and its graph is
symmetric about the straight line that goes through any extremum point orthogonally
to the $y$-axis.

By $U (y; s)$ we denote a solution of \eqref{eq:U} that satisfies the following
Cauchy data:
\begin{equation}
U (0; s) = 0 , \ \ U' (0; s) = s , \quad \mbox{where} \ s \geq 0 .
\label{eq:cd}
\end{equation}
To describe the behaviour of $U (y; s)$ we denote by $\tau_+ (s)$ and $\tau_- (s)$,
$s > 0$, the least positive and the largest negative root, respectively, of the
equation $2 \, \Omega (\tau) = s^2$. If this equation has no positive (negative)
root, we put $\tau_+ (s) = +\infty$ ($\tau_- (s) = -\infty$ respectively).
Furthermore, if $\omega (0) = 0$, then we put $\tau_\pm (0) = 0$, and if $\pm \omega
(0) > 0$, then $\tau_\pm (0) = 0$, whereas $\tau_\mp (0)$ is defined in the same way
as for $s > 0$. Considering
\begin{equation}
y_\pm (s) = \int_0^{\tau_\pm (s)} \frac{\D \tau}{\sqrt{s^2 - 2 \, \Omega (\tau)}} \,
, \label{eq:y_pm}
\end{equation}
we see that $y_+ (s)$ is finite if and only if $\tau_+ (s) < +\infty$ and $\omega
(\tau_+ (s)) > 0$, whereas the inequalities $\tau_- (s) > -\infty$ and $\omega
(\tau_- (s)) < 0$ are necessary and sufficient for finiteness of $y_- (s)$.

In terms of $y_\pm (s)$ the maximal interval, where the function $U$ given by the
implicit formula
\begin{equation}
y = \int_0^U \frac{\D \tau}{\sqrt{s^2 - 2 \Omega (\tau)}} 
\label{eq:Uim}
\end{equation}
increases strictly monotonically, is $(y_- (s), y_+ (s))$. Furthermore, if $y_-
(s)$ is finite, then $U' (y_- (s); s)$ vanishes and
\[ \tau_- (s) = U (y_- (s); s) = \min_{y \in \RR} U (y; s) > -\infty .
\]
Similarly, if $y_+ (s)$ is finite, then $U' (y_+ (s); s)$ vanishes and 
\[ \tau_+ (s) = U (y_+ (s); s) = \max_{y \in \RR} U (y; s) < +\infty .
\]
Otherwise, $\min$ and $\max$ must be changed to $\inf$ and $\sup$, respectively, in
these formulae.

Further properties of $U (y; s)$ given by the implicit equation \eqref{eq:Uim} on
$(y_- (s), y_+ (s))$ are as follows:

\noindent $\bullet$ If $y_+ (s) = +\infty$ and $y_- (s) = -\infty$, then $U (y; s)$
increases strictly monotonically for all $y \in \RR$.

\noindent $\bullet$ If $y_- (s) = -\infty$ and $y_+ (s) < +\infty$, then $U (y; s)$
attains its maximum at $y = y_+ (s)$ and decreases monotonically away from this
point in both directions of the $y$-axis.

\noindent $\bullet$ If $y_- (s) > -\infty$ and $y_+ (s) = +\infty$, then $U (y; s)$
attains its minimum at $y = y_- (s)$ and increases monotonically away from this
point in both directions of the $y$-axis.

\noindent $\bullet$ If both $y_+ (s)$ and $y_- (s)$ are finite, then $U (y; s)$ is
periodic; it attains one of its minima at $y = y_- (s)$ and one of its maxima at $y
= y_+ (s)$. Moreover, $U (y; s)$ increases strictly monotonically on $[y_- (s) , \,
y_+ (s)]$.

Second, we consider the problem
\begin{equation}
u'' + \omega (u) = 0 \ \ \mbox{on} \ (0, h) , \quad u (0) = 0 , \ \ u (h) = 1 ,
\label{eq:u}
\end{equation}
which involves one-dimensional versions of relations
\eqref{eq:lapp}--\eqref{eq:kcp}. It is clear that formula \eqref{eq:Uim} defines a
solution of problem \eqref{eq:u} on the interval $(0, h (s))$, where
\begin{equation}
h (s) = \int_0^1 \frac{\D \tau}{\sqrt{s^2 - 2 \, \Omega (\tau)}} \quad \mbox{and} \
s > s_0 = \max_{\tau \in [0, 1]} \sqrt{2 \, \Omega (\tau)} .
\label{eq:d}
\end{equation}
Furthermore, all monotonic solutions of problem \eqref{eq:u} have the form
\eqref{eq:Uim} on the interval $(0, h)$. This remains valid for $s = s_0$ with
\begin{equation}
 h = h_0 = \int_0^1 \frac{\D \tau}{\sqrt{s^2_0 - 2 \, \Omega (\tau)}} < \infty ,
\quad \mbox{that is}, \ h_0 = \lim_{s \to s_0} h (s) .
\label{eq:h_0}
\end{equation}
It is clear that $h_0 = +\infty$ when $\omega$ satisfies conditions (i) and $h_0$ is
finite otherwise; moreover, $h (s)$ is a strictly monotonically decreasing function
with values in $(0, h_0]$. 

Furthermore, the pair $(\psi, \eta)$ with $\psi = u (y) = U (y; s)$ and $\eta = h
(s)$ satisfies problem \eqref{eq:lapp}--\eqref{eq:bep} provided $s$ satisfies the
equation
\begin{equation}
{\cal R} (s) = r \, , \quad \mbox{where} \ {\cal R} (s) = [ s^2 - 2 \, \Omega (1) +
2 \, h (s) ] / 3 \, .
\label{eq:calR}
\end{equation}
This function has only one minimum, say $r_c > 0$, attained at some $s_c > s_0$.
Hence if $r \in (r_c, r_0)$, where
\[ r_0 = \lim_{s \to s_0 + 0} {\cal R} (s) = \frac{1}{3} \left[ s^2_0 -
2 \, \Omega (1) + 2 \, h_0 \right] ,
\]
then equation \eqref{eq:calR} has two solutions $s_+$ and $s_-$ such that $s_0 < s_+
< s_c < s_-$. By substituting $s_+$ and $s_-$ into \eqref{eq:Uim} and \eqref{eq:d},
one obtains the so-called {\it stream solutions} $(u_+, H_+)$ and $(u_-, H_-)$,
respectively. Indeed, these solutions satisfy the Bernoulli equation $u'_\pm (H_\pm)
+ 2 H_\pm = 3 r$ along with relations \eqref{eq:u}.

It should be mentioned that $s_-$ and the corresponding $H_-$ exist for all values
of $r$ greater than $r_c$, whereas $s_+$ and $H_+$ exist only when $r$ is less than
or equal to $r_0$; in the last case $s_+ = s_0$.

\subsubsection{Solutions with a single minimum}

Let $\omega$ satisfy conditions (ii), then $h_0 < \infty$ and $s_0 = 0$. Without
loss of generality, we consider $\omega$ as extended to $(-\infty, 0)$, and so by
$y^>$ we denote the largest zero of $\omega$ on $(-\infty, 0)$ and set $y^> =
-\infty$ when $\omega$ is positive throughout $(-\infty, 0)$. Putting $s^> = \sqrt{2
\, \Omega (y^>)}$, we see that the function $\tau_- (s)$ attains finite values on
$[0, s^>)$ and is continuous there; moreover, $y_- (s) \to -\infty$ as $s \to s^>$
provided $y^>$ is finite.

What was said in \S\,1.2 implies that for $s \in [0, s^>)$ the Cauchy problem with
data \eqref{eq:cd} has a solution such that
\[ U' (y_- (s); s) = 0 \quad \mbox{and} \quad U (2 y_- (s); s) = 0 .
\]
Now, putting
\begin{equation}
h_- (s) = h (s) - 2 y_- (s) ,
\label{eq:h_-}
\end{equation}
we see that the function
\begin{equation}
u_- (y; s) = U (y + 2 y_- (s); s)
\label{eq:u_-}
\end{equation}
solves problem \eqref{eq:u} on the interval $(0, h_- (s))$. Moreover, if $s$ is
determined from equation \eqref{eq:calR} with $r > r_0$, then $u_- (y; s)$ describes
a shear flow that has a counter-current near the bottom because $u_- (y; s)$ has a
single minimum at $y = - y_- (s)$.

For $\omega \equiv -2$ formulae \eqref{eq:cd} and \eqref{eq:u_-} take the form 
\[ U (y; s) = y^2 + s y \quad \mbox{and} \quad u_- (y; s) = y^2 - s y 
\]
respectively, whereas according to formula \eqref{eq:h_-} we have $h_- (s) = (s +
\sqrt{s^2 + 4}) / 2$ which is greater than $h_0 = 1$ for all $s > 0$; these
quantities are illustrated for $s=1$ in Figure~1.

\begin{figure}[t!]\centering
  \SetLabels
  \L (0.9*0.2) $y$\\
  \endSetLabels
  \leavevmode\AffixLabels{\includegraphics[width=80mm]{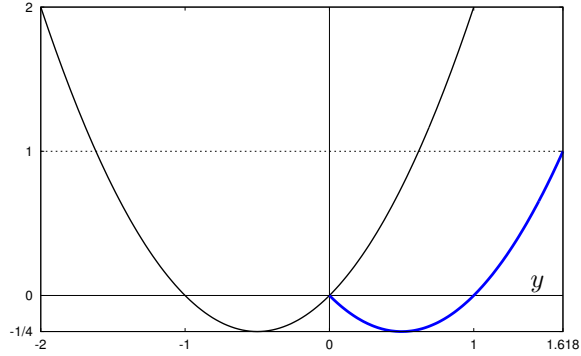}}
  \caption{$U (y; 1) = y^2 + y$ is plotted for $y \in (-2, 1)$ and $u_- (y; 1) = y^2 -
  y$ is plotted for $y \in (0, 1.618)$ because $h_- (1) = (1 + \sqrt{5}) / 2 \approx
  1.618$.}
  \label{fig:1}
\end{figure}

\subsubsection{Solutions with a single maximum}

Let  $\omega$ satisfy conditions (iii), and so $s_0 = \Omega (1)$ and $h_0 <
\infty$. Without loss of generality, we consider $\omega$ as extended to $(1,
+\infty)$, and so by $y^{<}$ we denote the least zero of $\omega$ on $(1, +\infty)$
and set $y^{<} = +\infty$ when $\omega$ is positive throughout $(1, +\infty)$.
Putting $s^{<} = \sqrt{2 \, \Omega (y^{<})}$, we see that the function $\tau_+ (s)$
attains finite values on $[s_0, s^{<})$ and is continuous there. It should be noted
that $y_+ (s) \to +\infty$ as $s \to s^{<}$ provided $y^<$ is finite.

What was said in \S\,1.2 implies that for $s \in [s_0, s^{<})$ the Cauchy problem
with data \eqref{eq:cd} has the solution for which
\[ U' (y_+ (s); s) = 0 \quad \mbox{and} \quad U (h (s) + 2 [y_+ (s) - h (s)]; s)
= 1 .
\]
Let us put
\begin{equation}
h_+ (s) = h (s) + 2 [y_+ (s) - h (s)] ,
\label{eq:h_+}
\end{equation}
then the function
\begin{equation}
u_+ (y; s) = U (y; s)
\label{eq:u_+}
\end{equation}
solves problem \eqref{eq:u} on the interval $(0, h_+ (s))$. Moreover, if $s$ is
found from equation \eqref{eq:calR} with $r > r_0$, then $u_+ (y; s)$ describes a
shear flow that has a counter-current near the free surface because $u_+ (y; s)$ has
a single maximum at $y = y_+ (s)$.

For $\omega \equiv 2$ both formulae \eqref{eq:cd} and \eqref{eq:u_+} give $U (y; s)
= u_+ (y; s) = -y^2 + s y$, whereas according to formula \eqref{eq:h_+} we have $h_+
(s) = (s + \sqrt{s^2 - 4}) / 2$ which is greater than $h_0 = 2$ for all $s > s_0 =
2$; these quantities are illustrated for $s=3$ in Figure~2.

\begin{figure}[t!]\centering
  \SetLabels
  \L (0.88*0.48) $y$\\
  \endSetLabels
  \leavevmode\AffixLabels{\includegraphics[width=45mm]{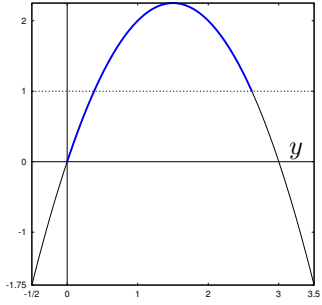}}
  \caption{$U (y; 3) = -y^2 + 3y$ is plotted for $y \in (-1/2, 7/2)$, and it coincides
  with $u_+ (y; 3)$ for $y \in (0, 2.618)$ (plotted bold) because $h_+ (3) = (3 +
  \sqrt{5}) / 2 \approx 2.618$.}
  \label{fig:2}
\end{figure}

\subsection{Main results}

Bounds for non-stream solutions of problem \eqref{eq:lapp}--\eqref{eq:bep} are
formulated in terms of solutions to problem \eqref{eq:u} and the characteristics
$r_c$, $H_-$ and $H_+$. As in the irrotational case (see \cite{KK0}), the depths
$H_-$ and $H_+$ of properly chosen supercritical and subcritical shear flows
respectively (they are also referred to as conjugate streams) serve as bounds for
$\hat{\eta}$ and $\check{\eta}$.

We begin with the following two theorems generalising Theorems~3.2 and 3.4 in
\cite{KK4}. The first of them asserts, in particular, that all free-surface profiles
subject to reasonable assumptions are located above the supercritical level.
Moreover, these theorems provide bounds for $\check \eta$ (the upper one), $\hat
\eta$ (the lower one) and $r$, which cannot be less than the critical value $r_c$.

\vspace{2mm}

\noindent {\bf Theorem 1.} {\it Let problem \eqref{eq:lapp}--\eqref{eq:bep} have a
non-stream solution $(\psi, \eta)$ such that $\psi \leq 1$ on $\bar D$. Then the
following two assertions are true.}

1. {\it If $\check \eta < h_0$, then $\psi (x, y) < \check U (y; \check s)$ in $\RR
\times (0, \check \eta)$ and $\check U (y; \check s)$ is given by formula
\eqref{eq:Uim} with $s = \check s$; here $\check s > s_0$ is such that $h (\check s)
= \check \eta$. Besides,} (A) $r \geq r_c$, (B) $H_- < \eta (x)$ {\it for all $x \in
\RR$, and if $r \leq r_0$, then also} (C) $\check \eta \leq H_+$. {\it Moreover,
the inequalities for $r$ and $\check \eta$ are strict provided the latter value is
attained at some point on the free surface.}

2. {\it Relations} (A)--(C) {\it are true when $\check \eta = h_0 \neq +\infty$.}

\vspace{2mm} 

\noindent {\bf Theorem 2.} {\it Let problem \eqref{eq:lapp}--\eqref{eq:bep} has a
non-stream solution $\left( \psi, \eta \right)$ such that $\psi \geq 0$ on $\bar D$.
Then the following two assertions are true.}

1. {\it If $\hat \eta < h_0$, then $\psi (x, y) > \hat U (y; \hat s)$ in $D$ and
$\hat U (y; \hat s)$ is given by formula \eqref{eq:Uim} with $s = \hat s$; here
$\hat s > s_0$ is such that $h (\hat s) = \hat \eta$. Moreover, the inequality
$\hat \eta \geq H_+$ holds provided $r \leq r_0$ and is strict when $\hat \eta$ is
attained at some point on the free surface.}

2. {\it The inequality $\hat \eta \geq H_+$ holds when $\hat \eta = h_0 \neq
+\infty$.}

\vspace{2mm}

In the next theorem that generalises Theorem~3.6 in \cite{KK4}, $\omega$ satisfies
conditions (ii) and the inequality satisfied by $\check{\eta}$ in Theorem~1 is
violated. In this case, an upper bound for $\psi$ is formulated in terms of the
family $(u_- (y; s), h_-(s))$ whose components depend on the parameter $s > s_0 = 0$
according to formulae \eqref{eq:u_-} and \eqref{eq:h_-} respectively. It is
essential that $u_-' (y; s)$ changes its sign on $(0, h_- (s))$ being negative near
the bottom, and so the flow described by $\psi$ also has a counter-current near the
bottom.

\vspace{2mm}

\noindent {\bf Theorem 3.} {\it Let $\omega$ satisfy conditions {\rm (ii)}. If
problem \eqref{eq:lapp}--\eqref{eq:bep} has a non-stream solu\-tion $(\psi, \eta)$
satisfying the inequalities $\psi \leq 1$ in $\bar D$ and $h_0 < \check \eta$, then
there exists a small $s_* > 0$ such that $h_0 < h_- (s_*) < \check \eta$ and $\psi
(x, y) < u_- (y; s_*)$ in $\RR \times (0, h_- (s_*))$.}

\vspace{2mm}

When $\omega$ satisfies conditions (iii) and the inequality imposed on $\hat \eta$
in Theorem~2 is violated we give a lower bound for a non-negative stream function
$\psi$ provided an extra condition is fulfilled for $\Omega$. This bound involves a
function characterised by formula \eqref{eq:Uim}. It is essential that the
derivative of this function changes its sign being negative near the free surface,
and so the flow described by $\psi$ also has a counter-current there.

\vspace{2mm}

\noindent {\bf Theorem 4.} {\it Let $\omega$ satisfy conditions {\rm (iii)} and
$\Omega (1) > 0$. If problem \eqref{eq:lapp}--\eqref{eq:bep} has a non-stream
solution $(\psi, \eta)$ for which $\psi \geq 0$ in $\bar D$ and $h_0 < \check \eta$,
then there exists $s^* > s_0$ such that $s^* - s_0$ is small, $h_0 < h_+ (s^*) < \check
\eta$ and the inequality $\psi (x, y) > u_+ (y; s_*)$ holds in $D$. Here $h_+ (s^*)$
and $u_+ (y; s^*)$ are given by formulae \eqref{eq:h_+} and \eqref{eq:u_+}
respectively.}

\vspace{2mm}

The last theorem generalises Theorem~3.8 in \cite{KK4}.

\section{Two lemmas}

In Lemmas 1 and 2, we analyse the asymptotic behaviour as $s \to s_0$ for the
functions defined by formulae \eqref{eq:h_-} and \eqref{eq:h_+} respectively. Here
and below one dot on top denotes the first derivative with respect to $s$.

\vspace{2mm}

\noindent {\bf Lemma 1.} {\it Let $\omega$ satisfy conditions {\rm (ii)}. Then the
following asymptotic formula holds
\begin{equation}\label{L1a}
h_- (s) = h (0) - \frac{s}{\omega (0)} + O (s^2) \quad \mbox{as} \ s \to 0 ,
\end{equation}
and so $\dot h_- (0) > 0$.  Moreover, if the function $h_-$  strictly increases on
$[0, s^{(0)}]$ for some $s^{(0)}$, then $u_- (y; s_1) > u_- (y; s_2)$ for all $y \in
(0, h_- (s_1)]$, where $s_1$ and $s_2$ are such that $0 \leq s_1 < s_2 \leq
s^{(0)}$.}

\vspace{2mm}

\noindent {\it Proof.} Let $a = \omega (0)$, and let $x$ be such that the equality
$a x = 2 \, \Omega (\tau)$ holds for small $\tau$. Hence $\tau_- (s) = s^2/a$, and
the change of variable gives
\[ y_-(s) = \int_0^{\tau_-} \frac{\D \tau}{\sqrt{s^2 - 2 \, \Omega (\tau)}} =
\int_0^{s^2/a} \frac{a \, \D x}{2 \, \omega (\tau (x)) \, \sqrt{s^2 - ax}} \, .
\]
Since $\omega (\tau (x)) = a + O (s^2)$ for small $s$ and $x$, we see that
\begin{equation}\label{L1b}
y_- (s) = \frac{s}{a} + O (s^3) \quad \mbox{as} \ s \to 0 .
\end{equation}
Furthermore, for sufficiently small $b$ we have
\[ h(s) = h (0) + \int_0^b \Bigg[ \frac{\D \tau}{\sqrt{s^2 - 2 \, \Omega (\tau)}} -
\frac{\D \tau}{\sqrt{-2\Omega (\tau)}} \Bigg] + O (s^2) \quad \mbox{as} \ s \to 0 .
\]
Using the same change of variable $a x = 2 \, \Omega (\tau)$, we write this as
follows:
\[ h (s) = h (0) + \int_0^{- a \tilde{b}/s^2} \frac{a}{2 \omega (\tau(x))} \Bigg[
\frac{\D x}{\sqrt{s^2 - a x}} - \frac{\D x}{\sqrt{-a x}} \Bigg] + O (s^2) \quad
\mbox{as} \ s \to 0 , 
\]
where $a \tilde{b} = 2 \Omega (b)$. Using again $\omega (\tau(x)) = a + O (x^2)$,
we see that
\begin{equation*}
h (s) = h (0) + \frac{1}{2} \int_0^{a\tilde{b}/s^2} \Bigg[ \frac{\D x}{\sqrt{s^2 - a
x}} - \frac{\D x}{\sqrt{-a x}} \Bigg] + O (s^2) = h (0) + \frac{s}{a} + O (s^2)
\quad \mbox{as} \ s \to 0 .
\end{equation*}
Since $h_- (s)$ is defined by formula (\ref{eq:h_-}), combining the last asymptotics
and (\ref{L1b}), we arrive at the required formula \eqref{L1a}.

To prove the second assertion we assume the contrary, that is,
\[ u_- (y^*, s_1) \leq u_- (y^*, s_2) \quad \mbox{for some} \ y^*\in (0,h_-(s_1)] .
\]
Diminishing $s_1$ and observing that $u_- (y; 0) > u_- (y; s_2)$ when $y \in (0, h_-
(0)]$, we conclude that there exists $\tilde s \in (0, s_1)$ and $\tilde y$ such
that $u_- (\tilde y; \tilde s) = u_- (\tilde y; s_2)$ and $u_- (y; \tilde s) \geq
u_- (y^*; s_2)$ on $(0, h_- (\tilde s)]$, which is impossible in view of the maximum
principle for non-negative functions.

The proof is complete.

\vspace{2mm}

\noindent {\bf Lemma 2.} {\it Let $\omega$ satisfy conditions {\rm (iii)}. Then the
following asymptotic formula holds}
\begin{equation}\label{L2a}
h_+(s) = h (0) + \frac{s^2 - s_0^2}{\omega (1)} + O (s -s_0) \quad as \ s \to s_0 .
\end{equation}

\vspace{2mm}

\noindent {\it Proof.} According to formula \eqref{eq:h_+}, we have to evaluate
\[ y_+ (s) - h (s) = \int_1^{\tau_+ (s)} \frac{\D \tau}{\sqrt{s^2 -2 \, \Omega (\tau)}}
\]
when $s$ is close to $s_0$. By changing variable to $x = 2 \, [\Omega (\tau) -
\Omega (1)] / b$, where $b = \omega (1)$ is positive, we obtain
\[ y_+ (s) - h (s) = \int_0^{x_+} \frac{b \, \D x}{2 \, \omega (\tau(x)) \sqrt{s^2 
- s_0^2 - b \, x}} = \frac{\sqrt{s^2-s_0^2}}{2b} \int_0^1 \frac{\D y}{\sqrt{1-y}} +
O (s^2-s_0^2) 
\]
as $s \to s_0$, where
\[ x_+ = \frac{s^2-s_0^2}{b} = \frac{2}{b} \int_1^{\tau_+ (s)} \omega(t) \, \D t . 
\]
This gives that
\begin{equation}\label{L2b}
y_+ (s) - h (s) = \frac{\sqrt{s^2-s_0^2}}{b} + O (s^2-s_0^2) \quad \mbox{as} \ s \to
s_0 .
\end{equation}
Furthermore, in
\[ h (s_0) -h (s) = \int_0^1 \frac{\D \tau}{\sqrt{\int_\tau^1 \omega (t) \, \D t}} 
- \int_0^1 \frac{\D \tau}{\sqrt{s^2-s_0^2 + \int_\tau^1 \omega (t) \D t}}.
\]
we change variable $\tau$ to
\[ y = \frac{2}{s^2-s_0^2} \int_\tau^1 \omega (t) \D t ,
\]
thus obtaining
\[ h (s_0) - h (s) = \frac{\sqrt{s^2-s_0^2}}{2b} \int_0^{\Omega(1)/(s^2-s_0^2)} 
\Bigg[ \frac{1}{\sqrt{y}} - \frac{1}{\sqrt{1+y}} \Bigg] \D y + O (s^2-s_0^2)
\]
as $s \to s_0$. Therefore,
\begin{equation*}
h (s_0) - h (s) = \frac{\sqrt{s^2-s_0^2}}{b}+O(s^2-s_0^2).
\end{equation*}
This formula and (\ref{L2b}) imply (\ref{L2a}), which completes the proof.

\section{Proof of main results}

Without loss of generality, we suppose that the vorticity distribution $\omega (t)$
is prescribed only on the range of $\psi$. Taking into account how Theorems 1--4 are
formulated, this range belongs to the half-axis $t \leq 1$ ($t \geq 0$) in
Theorems~1 and 3 (Theorems~2 and 4 respectively).

\subsection{Proof of Theorem 1}

First, let us consider the case when $\check \eta < h_0$, and so there exists
$\check s > s_0$ such that $h (\check s) = \check \eta$. Thus, the function $\check
U (y; \check s)$ solves problem \eqref{eq:u} on $(0, \check \eta)$. Moreover, this
formula defines $\check U (y; \check s)$ for all $y \geq 0$ provided $\omega (t)$ is
extended to $t > 1$ as a Lipschitz function for which the inequality $\check s^2 > 2
\max_{\tau \geq 0} \Omega (\tau)$ holds. For this purpose it is sufficient to extend
$\omega$ as a linear function to a small interval on the right of $t = 1$ and to put
$\omega \equiv 0$ farther right. Then we have
\[ \check U' (y; \check s) = \sqrt{\check s^2 - 2 \, \Omega (\check U (y; \check s))}
> 0 \quad \mbox{for all} \ y \geq 0 ,
\]
and so $\check U (y; \check s)$ is a monotonically increasing function of $y$ and
$\check U (y; \check s) > 1$ for $y > h_0$.

Let $U_\ell (y) = \check U (y + \ell; \check s)$ for $\ell \geq 0$, and so $U_\ell
(y) > 1$ on $[0, \check \eta]$ when $\ell > \check \eta$. Then $U_\ell - \psi > 0$
on $\RR \times [0, \check \eta]$ for $\ell > \check \eta$. Let us show that there is no
$\ell_0 > 0$ such that
\begin{equation}
\inf_{\RR \times [0, \check \eta]} (U_{\ell_0} - \psi) = 0 . \label{inf}
\end{equation}
Assuming that such $\ell_0$ exists, we see that $U_{\ell_0} - \psi$ attains values
separated from zero on both sides of the strip $\RR \times [0, \check \eta]$. Since
$\psi$ satisfies conditions \eqref{bound}, there exist positive $\epsilon$ and $\delta$
such that
\begin{equation}
U_{\ell_0} (y) - \psi (x,y) \geq \epsilon \quad \mbox{when} \ (x,y) \in (\RR \times
[0, \delta]) \cup (\RR \times [\check \eta - \delta, \check \eta]) .
\label{e_d}
\end{equation}
Therefore, \eqref{inf} holds when either $U_{\ell_0} (y_0) - \psi (x_0, y_0) = 0$
for some $(x_0, y_0) \in \RR \times (0, \check \eta)$ or there exists a sequence
$\{(x_k, y_k)\}_{k=1}^\infty \subset \RR \times (\delta, \check \eta - \delta)$ such
that
\begin{equation}
U_{\ell_0} (y_k) - \psi (x_k, y_k) \to 0 \quad \mbox{as} \ k \to \infty .
\label{seq}
\end{equation}

The first of these options is impossible. Indeed, the elliptic equation
\begin{equation}
\nabla^2 (U_{\ell_0} - \psi) + (U_{\ell_0} - \psi) \int_0^1 \omega' (t [U_{\ell_0} -
\psi]) \, \D t = 0 \quad \mbox{holds in} \ \RR \times (0, \check \eta) .
\label{ell}
\end{equation}
Then the maximum principle (see \cite{GNN}, p.~212) guarantees that the non-negative
function $U_{\ell_0} - \psi$ cannot vanish at $(x_0, y_0) \in \RR \times (0, \check
\eta)$ because otherwise $\psi$ must coincide with $U_{\ell_0}$ everywhere.

In order to show that the second option \eqref{seq} is also impossible we apply
Harnack's inequality (see Corollary~8.21 in \cite{GT}) to the last equation (indeed,
$U_{\ell_0} - \psi$ is positive in $\RR \times [0, \check \eta]$). Therefore, there
exists $C > 0$ such that
\[ \sup_{(x,y) \in B_\rho (x_k, \check \eta/2)} \big[ U_{\ell_0} (y)- \psi (x,y) \big] 
\leq C \inf_{(x,y) \in B_\rho (x_k, \check \eta/2)} \big[ U_{\ell_0} (y) - \psi
(x,y) \big]
\]
in every circle $B_\rho (x_k, \check \eta/2)$, $k=1,2,\dots$, with $\rho = (\check
\eta - \delta) / 2$. Then \eqref{seq} yields that the supremum on the left-hand
side is arbitrarily small provided $k$ is sufficiently large, but this is
incompatible with \eqref{e_d}.

The obtained contradictions show that there is no $\ell_0 > 0$ such that \eqref{inf}
is true. Letting $\ell_0 \to 0$, we see that $\check U (y; \check s) - \psi (x,y)$
is non-negative on $\RR \times [0, \check \eta]$ and vanishes when $y = 0$. Since
this function satisfies equation \eqref{ell} with $\ell_0 = 0$, the maximum
principle implies that it does not vanish at inner points of the strip $\RR \times
(0, \check \eta)$, and so $\check U (y; \check s) - \psi (x, y) > 0$ throughout this
strip. Thus, the first inequality of assertion~1 is proved.

Let us show that relations (A)--(C) hold. First, we suppose that there exists $x_0
\in \RR$ such that $\eta (x_0) = \check \eta$ (it is clear that $y = \eta (x)$ is
tangent to $y = \check \eta$ at $x_0$). Then $\check U (\check \eta; \check s) - \psi
(x_0, \check \eta) = 0$ because both terms are equal to one at this point. Now, it
follows from Hopf's lemma (see \cite{GNN}, p.~212) that
\[ \big[ \check U' (y; \check s) - \psi_y (x,y) \big]_{(x,y)=(x_0,\check \eta)} < 0 \, .
\]
Taking into account Bernoulli's equation at $(x_0, \check \eta)$, that is, $\psi_y
(x_0, \check \eta) = \sqrt{3 r - 2 \check \eta}$, we obtain that $\check U' (\check
\eta; \check s) < \sqrt{3 r - 2 \check \eta}$ which is equivalent to
\[ \check s^2 - 2 \Omega (1) < 3 r - 2 \check \eta , \quad \mbox{and so} \quad
{\cal R} (\check s) < r \ \mbox{according to \eqref{eq:calR}}.
\]
The last inequality yields that relations (A)--(C) of assertion~1 are true and
inequalities (A) and (C) are strict in this case.

In the alternative case, we have that $\eta (x) > \check \eta$ for all $x \in \RR$
and there exists a sequence
\[ \{ x_k \}_{k=1}^\infty \subset \RR \quad \mbox{such that} \ \eta (x_k) \to \check
\eta \ \mbox{as} \ k \to \infty .
\]
Let us put $\check u (x,y) = \check U (y; \check s) - \psi (x, y)$ and consider the
behaviour of $\check u (x_k, \check \eta)$ and the derivatives of $\check u$ at
$(x_k, \check \eta)$ as $k \to \infty$.

Since $\check u (x_k, \check \eta) = \psi (x_k, \eta (x_k)) - \psi (x_k, \check
\eta)$, we see that this difference, say $\epsilon_k \geq 0$, tends to zero as $\ k
\to \infty$. Moreover,
\begin{equation} 
\epsilon_k \quad \mbox{and} \quad \eta (x_k) - \check \eta \quad \mbox{tend to zero
as} \ k \to \infty .
\label{eq:11}
\end{equation}
Let us prove that $\check u_x (x_k, \check \eta)$ also tends to zero as $k \to
\infty$. Indeed, we have for $t > 0$:
\[ \check u (x_k \pm t, \check \eta) = \check u (x_k, \check \eta) \pm t \, \check u_x 
(x_k, \check \eta) + \alpha (t) \quad \mbox{and} \ \alpha (t) = o (t) \ \mbox{as} \
t \to 0 .
\]
By rearrangement we obtain that $\pm \check u_x (x_k, \check \eta) = t^{-1} \left[
\check u (x_k \pm t, \check \eta) - \epsilon_k \right] + t^{-1} \alpha (t)$, where
the first term in the squire brackets is positive. Therefore,
\[ \mp \check u_x (x_k, \check \eta) \leq t^{-1} \epsilon_k + t^{-1} \alpha (t) ,
\quad \mbox{and the last term is} \ o (1) \ \mbox{as} \ t \to 0 .
\]
Hence $|\check u_x (x_k, \check \eta)|$ is less than arbitrarily small $\delta > 0$
provided $k$ is sufficiently large. First, let $t$ be small enough to guarantee that
$t^{-1} \alpha (t) < \delta / 2$. Fixing this $t$, we use \eqref{eq:11} and take $k$
such that $t^{-1} \epsilon_k < \delta / 2$. This implies that $|\check u_x (x_k,
\check \eta)| < \delta$, thus completing the proof that $\check u_x (x_k, \check
\eta) \to 0$ as $k \to \infty$.

According to the definition of $\check u$, this means that $\psi_x (x_k, \check
\eta) \to 0$ as $k \to \infty$. The next step is to show that
\begin{equation}
\underset{k \to \infty}{\rm lim\,sup} \ \check u_y (x_k, \check \eta) \leq 0 \, .
\label{eq:12}
\end{equation}
For $t > 0$ we have
\[ \check u (x_k, \check \eta - t) = \check u (x_k, \check \eta) - t \, \check u_y
(x_k, \check \eta) + \beta (t) \quad \mbox{and} \ \beta (t) = o (t) \ \mbox{as} \
t \to 0 .
\]
In the same way as above we obtain that
\[ \check u_y (x_k, \check \eta) \leq t^{-1} \epsilon_k + t^{-1} \beta (t) ,
\quad \mbox{where the last term is} \ o (1) \ \mbox{as} \ t \to 0 .
\]
It is clear that this implies \eqref{eq:12}. Then, taking a subsequence if necessary
and using Bernoulli's equations for $\check U$ and $\psi$, we let $k \to \infty$
and conclude that
\begin{equation}
\check s^2 - 2 \, \Omega (1) \leq 3 r - 2 \, \check \eta . \label{check_s}
\end{equation}
By rearranging this inequality, all three relations (A)--(C) follow.

Let us turn to assertion 2 when $\check \eta = h_0 \neq +\infty$. By $\{ h_j
\}_1^\infty$ we denote a sequence such that the inequalities $h (s_c) < h_j < h_0$
hold for its elements (see the line next to \eqref{eq:calR} for the definition of
$s_c$), and $h_j \to h_0$ as $j \to \infty$. Then the function inverse to
\eqref{eq:d} gives $s_j$ for which $h (s_j) = h_j$, and so $s_j \to s_0$ as $j \to
\infty$. Moreover, we have the stream solution $(U (y; s_j), h_j)$ with the first
component defined on $[0, h_j]$ by formula \eqref{eq:Uim}. Thus, each function of
the sequence $u_j (x, y) = U (y; s_j) - \psi (x, y)$, $j=1,2,\dots$, is defined on
$\RR \times [0, h_j]$.

By the definition of $\check \eta$ there exists a sequence $\{ x_j \}_1^\infty
\subset \RR$ (it is possible that all its elements coincide) such that $\eta (x_j) -
h_j \to 0$ as $j \to \infty$. Then considerations similar to those above yield that
\[ \partial_x u_j (x_j, h_j) \to 0 \ \mbox{as} \ j \to \infty \quad \mbox{and} \quad 
\underset{j \to \infty}{\rm lim\,sup} \ \partial_y u_j (x_j, h_j) \leq 0 ,
\]
which leads to the inequality
\[ s_0^2 - 2 \, \Omega (1) \leq 3 r - 2 \, \check \eta .
\]
instead of \eqref{check_s}. Since $\check \eta = h_0$ this inequality, gives all three
required assertions. The proof is complete.

\subsection{Proof of Theorem 2}

At its initial stage the proof of this theorem is similar to that of Theorem~1.
Namely, we consider the case when $\hat \eta < h_0$ first. Since there exists $\hat
s > s_0$ such that $h (\hat s) = \hat \eta$, the function $\hat U (y; \hat s)$
given by formula \eqref{eq:Uim} with $s = \hat s$ solves problem \eqref{eq:u} on
$(0, \hat \eta)$. Moreover, the same formula defines this function for all $y \leq
\hat \eta$ provided $\omega (t)$ is extended to $t < 0$ as a Lipschitz function for
which the inequality $\hat s^2 > 2 \max \Omega (\tau)$ holds. As in the proof of
Theorem~1, it is sufficient to extend $\omega$ as a linear function to a small
interval on the left of $t = 0$ and to put $\omega \equiv 0$ farther left. Then we
have
\[ \hat U' (y; \hat s) = \sqrt{\hat s^2 - 2 \, \Omega (\hat U (y; \hat s))}
> 0 \quad \mbox{for all} \ y \leq \hat \eta ,
\]
and so $\hat U (y; \hat s)$ is a monotonically increasing function of $y$ and $\hat
U (y; \hat s) < 0$ for $y < 0$.

Let $U_\ell (y) = \hat U (y - \ell; \hat s)$ for $\ell \geq 0$, and so $U_\ell (y) <
0$ on $[0, \hat \eta]$ when $\ell > \hat \eta$. Then $U_\ell - \psi < 0$ on $\bar D$
for $\ell > \hat \eta$. Let us show that there is no $\ell_0 > 0$ such that
\begin{equation}
\sup_{\bar D} \, (U_{\ell_0} - \psi) = 0 . \label{sup}
\end{equation}
Assuming that such $\ell_0$ exists, we see that $U_{\ell_0} - \psi$ attains values
separated from zero on both sides of $\bar D$.  Since $\psi$ satisfies conditions
\eqref{bound}, there exist positive $\epsilon$ and $\delta$ such that
\begin{equation}
U_{\ell_0} (y) - \psi (x,y) \geq \epsilon \quad \mbox{when} \ (x,y) \in (\RR \times
[0, \delta]) \cup \{ x \in \RR , y \in [\eta (x) - \delta, \eta (x)] \} .
\label{e_d'}
\end{equation} 
Therefore, \eqref{sup} holds when either $U_{\ell_0} (y_0) - \psi (x_0, y_0) = 0$
for some $(x_0, y_0) \in \bar D$ or there exists a sequence $\{(x_k,
y_k)\}_{k=1}^\infty \subset \{ x \in \RR , y \in ( \delta, \eta (x) - \delta ) \}$
such that
\begin{equation}
U_{\ell_0} (y_k) - \psi (x_k, y_k) \to 0 \quad \mbox{as} \ k \to \infty .
\label{seq'}
\end{equation}
The same methods as in the proof of Theorem~1 demonstrate that either of the options
\eqref{e_d'} and \eqref{seq'} leads to a contradiction, and so there is no $\ell_0 >
0$ such that \eqref{sup} is true. Letting $\ell_0 \to 0$, we see that $\hat U (y;
\hat s) - \psi (x,y)$ is non-positive on $\bar D$ and vanishes when $y = 0$. Since
this function satisfies equation \eqref{ell} with $\ell_0 = 0$, the maximum
principle implies that it does not vanish at inner points of the strip $D$, and so
$\hat U (y; \hat s) - \psi (x, y) < 0$ throughout this strip. Thus, the first
inequality of assertion~1 is proved.

To show the inequality $\hat \eta \geq H_+$ we argue by analogy with the proof of
Theorem~1. First, we suppose that there exists $x_0 \in \RR$ such that $\eta (x_0) =
\hat \eta$ (it is clear that $y = \eta (x)$ is tangent to $y = \hat \eta$ at $x_0$).
Then $\check U (\hat \eta; \hat s) - \psi (x_0, \hat \eta) = 0$ because both terms
are equal to one at this point. Now, it follows from Hopf's lemma (see \cite{GNN},
p.~212) that
\[ \big[ \hat U' (y; \hat s) - \psi_y (x,y) \big]_{(x,y)=(x_0,\hat \eta)} > 0 \, .
\]
Taking into account Bernoulli's equation at $(x_0, \check \eta)$, we obtain that
\begin{equation}
\hat U' (\hat \eta; \hat s) > \sqrt{3 r - 2 \hat \eta} \label{17oct}
\end{equation}
because $\psi_y (x_0, \hat \eta) \geq 0$. Indeed, $\psi (x_0, \hat \eta) = 1$ and
$\psi \leq \hat U$ in a neighbourhood of $(x_0, \hat \eta)$, whereas $\hat U \leq
1$. A consequence of \eqref{17oct} is that the required inequality holds and is
strict.

In the alternative case, that is, when $\eta (x) < \hat \eta$ for all $x \in \RR$
and there exists a sequence $\{ x_k \}_{k=1}^\infty \subset \RR$ such that $\eta
(x_k) \to \hat \eta$ as $k \to \infty$, the considerations applied for proving the
corresponding part of Theorem~1 must be used with necessary amendments, thus leading
to the required inequality which is non-strict. Also, this remark concerns the proof
of assertion~2.

\subsection{Proof of Theorem 3}

Since $h_0 < \check \eta$, the function $\psi$ should be compared with a more
sophisticated family of test functions than $U_\ell (y) = \check U (y + \ell; \check
s)$, $\ell \geq 0$, used in the proof of Theorem~1. To define this family, say $v
(y; \lambda)$ depending on the parameter $\lambda \in [0, \Lambda]$ with $\Lambda$
to be described later, we, as in the proof of Theorem~1, extend $\omega (t)$ as a
linear function to a small interval on the right of $t = 1$ and put $\omega \equiv
0$ farther right. Thus, we obtain a Lipschitz function such that the inequality
$s_*^2 > 2 \max_{\tau \geq 0} \Omega (\tau)$ holds for $s_* > s_0 = 0$ which is so
small that $h_0 < h_- (s_*) < \check \eta$ and the function $u_- (y; s)$ is
well-defined for $s \in [0, s_*]$; see formulae \eqref{eq:h_-} and \eqref{eq:u_-}
for the definitions of $h_- (s)$ and $u_- (y; s)$ respectively.

Let us recall the properties of $u_- (y; s_*)$ essential for constructing the family
$v (y; \lambda)$ that depends on $(y; \lambda) \in [0, d_*] \times [0, \Lambda]$
continuously; $d_* = h_- (s_*)$. The function $u_- (y; s_*)$ solves problem
\eqref{eq:u} on the interval $(0, d_*)$; moreover, it attains its single negative
minimum at $- y_- (s_*)$ (see fig.~1). Furthermore, to apply considerations used in
the proof of Theorem~1 the following properties are required:

\vspace{1mm}

(I) $v'' + \omega (v) = 0$ on $(0, d_*)$ for $\lambda \in [0, \Lambda]$; (II) $v (y;
0) = u_- (y; s_*)$;

(III) $v (0; \lambda) > 0$ and $v (d_*; \lambda) > 1$ for $\lambda \in (0,
\Lambda)$; (IV) $v (y; \Lambda) > 1$ for all $y \in [0, d_*]$.

\vspace{1mm}

To construct $v (y; \lambda)$ we first consider $\lambda \in [0, s_*]$ and put
\begin{equation}
v (y; \lambda) = u_- (y - c \lambda; s_* - \lambda) \ \ \mbox{with} \ c \in (0, - [2
\, \omega (0)]^{-1}) \label{c} .
\end{equation}
Then properties (I), (II) and the first inequality (III) hold by the definition of
this function. In order to show that the second inequality is true we write
\begin{eqnarray*}
&& d_* = h_- (s_*) = h_- (s_* - \lambda) + \lambda h_-' (s_* - \lambda) + O
(\lambda^2) \nonumber \\ && \ \ \ \ = h_- (s_* - \lambda) - \lambda [\omega
(0)]^{-1} + O (s_* \lambda) \ \ \mbox{as} \ s_*, \lambda \to 0 .
\end{eqnarray*}
Here the second equality is a consequence of Lemma~1. Since $u_- (h_- (s_* -
\lambda) ; s_* - \lambda)$ is equal to one, we have
\begin{eqnarray*}
&& v (d_*; \lambda) = u_- ( h_- (s_* - \lambda) - \lambda [\omega (0)]^{-1} - c
\lambda + O (s_* \lambda) ; s_* - \lambda ) \\ && \ \ \ \ \ \ \ \ \ \ \ = 1 -
\lambda \{ [\omega (0)]^{-1} + c \} u_-' ( h_- (s_* - \lambda); s_* - \lambda ) + O
(s_* \lambda) \ \ \mbox{as} \ s_*, \lambda \to 0 .
\end{eqnarray*}
According to the definition of $c$, the expression in braces is negative, whereas
\[ u_-' ( h_- (s_* - \lambda); s_* - \lambda ) = \sqrt{s_*^2 - 2 \, \Omega (1)} > 0 . \]
Therefore, the second inequality (III) is true for $\lambda \in [0, s_*]$ provided
$s_*$ is sufficiently small.

The next step is to define $v (y; \lambda)$ for $\lambda \in [s_*, s_>]$ with $s_>$
such that $s_> - s_* > 0$ is small. Let $\lambda \in [s_*, s_>]$ and $V_\lambda (y)$
be given implicitly for $y \in [0, \infty)$ as follows:
\[ y = \int_{\lambda - s_*}^{V_\lambda} \frac{\D \tau}{\sqrt{2 \, [\Omega 
(\lambda - s_>) - \Omega (\tau)}} \, .
\]
Since $\omega$ satisfies conditions (ii) (in particular, $\omega (0) < 0$) and is
extended to the half-axis $\{ t \geq 1 \}$ so that the inequality $\Omega (\tau)
\leq \Omega (1) / 2$ holds for $\tau \geq 1$, we see that $V_\lambda (y)$
monotonically increases from $\lambda - s_*$ to infinity and $V_\lambda' (0) = 0$.
The last equality allows us to consider $V_\lambda (y)$ as an even function on the
whole real axis $\RR$.

Putting $v (y; \lambda) = V_\lambda (y - c \, s_>)$ with the same $c$ as in formula
\eqref{c}, we obtain the function $v (y; \lambda)$ continuous for $(y; \lambda) \in
[0, d_*] \times [0, s_>]$ for which properties (I)--(III) hold; in particular, the
second inequality (III) for $\lambda \in [s_*, s_>]$ is a consequence of preceding
considerations and the assumption that $s_> - s_*$ is small.

Finally, let $\Lambda = d_* + s_>$ and $v (y; \lambda) = v (y + \lambda - s_>; s_>)$
for $\lambda \in [s_>, \Lambda]$. For these values of $\lambda$ the second
inequality (III) follows by monotonicity from the same inequality for $\lambda \in
[s_*, s_>]$. Moreover, property (IV) is also a consequence of monotonicity and the
second inequality (III).

Having the family $v (y; \lambda)$, we proceed in the same way as in the proof of
Theorem~1. In view of property (IV) we have that
\[ v (y; \Lambda) - \psi (x, y) > 0 \quad \mbox{for all} \ (x, y) \in \RR \times 
[0, d_*] .
\]
Let us show that there is no $\lambda_0 \in (0, \Lambda)$ such that
\begin{equation}
\inf_{\RR \times [0, d_*]} \{ v (y; \lambda_0) - \psi (x, y) \} = 0 . \label{inf*}
\end{equation}
If such $\lambda_0$ exists, then inequalities (III) guarantee that $v (y; \lambda_0) -
\psi (x, y)$ attains values separated from zero on both sides of the strip $\RR
\times [0, d_*]$. Since $\psi$ satisfies conditions \eqref{bound}, there exist
positive $\epsilon$ and $\delta$ such that
\begin{equation*}
v (y; \lambda_0) - \psi (x,y) \geq \epsilon \quad \mbox{when} \ (x,y) \in (\RR
\times [0, \delta]) \cup (\RR \times [d_* - \delta, d_*]) .
\end{equation*}
Therefore, \eqref{inf*} holds when either 
\[ v (y_0; \lambda_0) - \psi (x_0, y_0) = 0 \ \ \mbox{for some} \ (x_0, y_0) \in 
\RR \times (0, d_*)
\]
or there exists a sequence $\{(x_k, y_k)\}_{k=1}^\infty \subset \RR \times (\delta,
d_* - \delta)$ such that
\begin{equation*}
v (y_k; \lambda_0) - \psi (x_k, y_k) \to 0 \quad \mbox{as} \ k \to \infty .
\end{equation*}
Literally repeating considerations in the proof of Theorem 1, one demonstrates that
both these options are impossible and the limit function $v (y; 0) - \psi (x,y)$ as
$\lambda \to 0$ is non-negative on $\RR \times [0, d_*]$ and vanishes when $y = 0$.

We recall that $v (y; 0) = u_- (y; s_*)$ by property (II). Since this function
satisfies property (I), we obtain that 
\[ \nabla^2 [u_- (\cdot; s_*) - \psi] + [u_- (\cdot; s_*) - \psi] \int_0^1 \omega'
(t [u_- (\cdot; s_*) - \psi]) \, \D t = 0 \ \ \mbox{in} \ \RR \times (0, d_*) .
\]
Then the maximum principle implies that $u_- (\cdot; s_*) - \psi$ does not vanish at
inner points of this strip, that is, $u_- (y; s_*) - \psi (x, y) > 0$ there.
Recollecting that $d_* = h_- (s_*)$, we arrive at the theorem's assertion.

\subsection{Proof of Theorem 4}

Since $\omega$ satisfies conditions (iii) and $\Omega (1) > 0$, we have that $s_0 =
\sqrt{2 \, \Omega (1)} > 0$. Since $h_0 < \check \eta$, a more sophisticated family of
test functions is required than that used in the proof of Theorem~2 (cf. the proof
of Theorem~3). To construct it we extend $\omega (t)$ in the same way as in the
proof of Theorem~2, that is, as a linear function to a small interval on the left of
$t = 0$ and put $\omega \equiv 0$ farther left so that the inequality $s_0^2 - 2 \,
\Omega (\tau) > s_0^2 / 2$ holds for $\tau \leq 0$. This implies that for every $s$
in a vicinity of $s_0$ the function $U (y; s)$ monotonically increases on the
half-axis $(-\infty, y_+ (s)]$ from $-\infty$ to the maximum value $\tau_+ (s) > 1$
and the graph of $U$ is symmetric about the vertical line through $y_+ (s)$.

Let $s^* > s_0$ be such that $s^* - s_0$ is so small that $U (y; s^*)$ is
well-defined, and $h_0 < d^* < \check \eta$; here and below $d^* = h_+ (s^*)$. Let us
consider $u_+ (y; s^*)$ that solves problem \eqref{eq:u} on the interval $(0, d^*)$;
this function coincides with $U (y; s^*)$ there. It should be noted that $u_+ (y;
s^*) > 1$ when $y \in (h (s^*), d^*)$; moreover, this function monotonically
increases from zero to its maximum value $\tau_+ (s^*) > 1$ on $[0, y_+ (s^*)]$ and
monotonically decreases from $\tau_+ (s^*)$ to one on $[y_+ (s^*), d^*]$.

Now we construct a family of test functions, say $w (y; \theta)$, continuously
depending on $(y; \theta) \in [0, \hat \eta] \times [0, \Theta]$ with $\Theta$ to be
described later. The following properties are analogous to those in the proof of
Theorem~3:

\vspace{1mm}

(I) $w'' + \omega (w) = 0$ on $(0, \hat \eta)$ for $\theta \in [0, \Theta]$; (II) $w
(y; 0) = u_+ (y; s^*)$;

(III) $w (0; \theta) < 0$ and $w (y; \theta) < 1$ for $\theta \in (0, \Theta)$ and $y
\in [d^*, \hat \eta]$;

(IV) $w (y; \Theta) < 0$ for all $y \in [0, \hat \eta]$.

\vspace{1mm}

First, we consider $\theta \in [0, s^* - s_0]$ and put
\begin{equation}
w (y; \theta) = u_+ (y - c \theta; s^* - \theta) \ \ \mbox{with} \ c \in (0, s_0
[\omega (1)]^{-1}) \label{c*} \, .
\end{equation}
This definition guarantees that properties (I), (II) and the first inequality (III)
hold for these values of $\theta$.

Let us show that the second inequality (III) is a consequence of $w (d^*; \theta) <
1$, for which purpose we check that $u_+ (y - c \, \theta; s^* - \theta)$
monotonically decreases when $y$ is greater than $d^*$. Since $u_+ (y; s)$ has this
property for $y > y_+ (s)$, it is sufficient to establish that $d^* - c \, \theta
> y_+ (s^* - \theta)$. Indeed, we have
\[ d^* - c \, \theta = h_+ (s^*) - c \, \theta = h_+ (s^* - \theta) - c \, \theta +
\theta \dot h_+ (s^* - \theta) + O (\theta^2) \quad \mbox{as} \ \theta \to 0 ,
\]
where $\dot h_+ (s^* - \theta) = 2 \, s_0 / \omega (1) + O (s^* - s_0 - \theta)$ in
view of Lemma~2. Combining this and the definition of $c$, we see that $\theta \dot
h_+ (s^* - \theta) - c \, \theta > 0$ provided $s^* - s_0$ is sufficiently small.
Furthermore, $h_+ (s^* - \theta) = y_+ (s^* - \theta) + y_+ (s^* - \theta) - h (s^*
- \theta)$, where the difference is non-negative, which together with the previous
inequality yields the required one.

To evaluate $w (d^*; \theta)$, we write
\begin{eqnarray*}
&& \!\!\!\!\!\!\!\!\!\!\!\!\!\! w (d^*; \theta) = u_+ (h_+ (s^* - \theta + \theta) -
c \, \theta; s^* - \theta) \\ && \ \ \ \ = u_+ (h_+ (s^* - \theta) + \theta \dot
h_+ (s^* - \theta) - c \, \theta + O (\theta^2); s^* - \theta) \quad \mbox{as} \
\theta \to 0 .
\end{eqnarray*}
Using Lemma 2 again, we obtain that
\begin{eqnarray*}
&& \!\!\!\!\!\!\!\!\!\!\!\!\!\! w (d^*; \theta) = u_+ (h_+ (s^* - \theta; s^* -
\theta) + [2 \, s_0 / \omega (1) - c] \theta + O ((s^* - s_0) \theta)) \\ && \ \ \ \
= 1 + u'_+ (h_+ (s^* - \theta); s^* - \theta) \left\{  [2 \, s_0 / \omega (1) - c]
\theta + O ((s^* - s_0) \theta) \right\} \\ && \ \ \ \ + \, 2^{-1} u''_+ (h_+ (s^* -
\theta); s^* - \theta) \left\{ [2 \, s_0 / \omega (1) - c] \theta + O ((s^* - s_0)
\theta) \right\}^2 \\ && \ \ \ \ + \, O \left( (s^* - s_0)^3 \right) \quad
\mbox{as} \ s^* - s_0 \to 0 ,
\end{eqnarray*}
because $u_+ (h_+ (s^* - \theta); s^* - \theta) = 1$. Let us show that the second
and third terms on the right-hand side are negative provided $s^* - s_0$ is
sufficiently small. Indeed, $2 \, s_0 / \omega (1) - c > 0$ by the definition of
$c$, and so the expression in braces is positive for such values of $s^*$ and
$\theta$. Therefore, it remains to investigate the behaviour of
\[ u'_+ (h_+ (s^* - \theta); s^* - \theta) \ \ \mbox{and} \ \ u''_+ (h_+ (s^* -
\theta); s^* - \theta)
\]
for these values. Since $\omega$ satisfies conditions (iii), $u'_+ (h_0; s_0) = 0$
which yields that
\[ u'_+ (h_+ (s^* - \theta); s^* - \theta) = u''_+ (h_0; s_0) [ (s^* - \theta)^2 -
s_0^2 ] / \omega (1) + O \left( (s^* - s_0)^2 \right) \ \ \mbox{as} \ s^* - s_0 \to
0 .
\]
This implies that
\[ u'_+ (h_+ (s^* - \theta); s^* - \theta) = s_0^2 - (s^* - \theta)^2 + O \left(
(s^* - s_0)^2 \right) \ \ \mbox{as} \ s^* - s_0 \to 0 ,
\]
in view of the equality $u''_+ (h_0; s_0) = - \omega (1)$, and so $u'_+ (h_+ (s^* -
\theta); s^* - \theta)$ is either negative or equal to $O \left( (s^* - s_0)^2
\right)$ when $s^*$ is sufficiently close to $s_0$ and $\theta \in [0, s^* - s_0]$.
Besides, the same equality gives that
\[ u''_+ (h_+ (s^* - \theta); s^* - \theta) = - \omega (1) + O (s^* - s_0 - \theta)
\ \ \mbox{as} \ s^* - s_0 \to 0 ,
\]
and the right-hand side is negative provided $s^*$ and $\theta$ have the same
properties. This is a consequence of $\omega (1) > 0$ which is included in
conditions (iii).

Using these facts in the last expression for $w (d^*; \theta)$, we see that it is
less than one for $\theta \in [0, s^* - s_0]$ provided $s^*$ is sufficiently close
to $s_0$.

The next step is to define $w (y; \theta)$ for $\theta \in [s^*, s^>]$ with $s^>$
such that $s^> - s^* > 0$ is small. In this case, formula \eqref{eq:Uim} defines the
function $U (y; s)$ for $y \in (-\infty, y_+ (1))$ and $s \in [s^*, s^>]$. This
allows us to put
\begin{equation} 
w (y; \theta) = U (y - c \, (s^* - s_0); s^* - \theta) \quad \mbox{for} \ \theta \in
[s^* - s_0, s^> - s^*] \label{c*'} \, ,
\end{equation}
where $c$ is the same as in \eqref{c*}. It is clear that property (I) is fulfilled
and both inequalities (III) hold because they are true for $\theta = s^* - s_0$ and
$s^> - s_0$ is small. Moreover, $w (y; \theta) < 1$ for $\theta$ described in
\eqref{c*'} and all $y \in \RR$ since 
\[ \sup_{y \in \RR} w (y; \theta) = \tau_+ (s^* - \theta) < 1 .
\]
Finally, the continuity follows from the fact that it holds for $\theta = s^* - s_0$
which one verifies directly.

Let $\Theta = \hat \eta - c \, (s^* - s_0) + s^* - s^> + \delta$, where $\delta > 0$
is sufficiently small. Putting
\[ w (y; \theta) = U (y - c \, (s^* - s_0) - \theta + s^* - s^>; s^>) \quad \mbox{for}
\ \theta \in [s^> - s^*, \Theta] ,
\]
we see that properties (I) and (III) are fulfilled by continuity, and so it remains
to check property (IV). Indeed, it follows from continuity and the fact that
\[  w (d^*; \Theta) = U (- \delta; s^>) < 0 .
\]

Having the family $w (y; \theta)$, we proceed in the same way as in the proof of
Theorem~2. In view of property (IV) we have that $w (y; \Theta) - \psi (x, y) < 0$
on $\bar D$. Let us assume that there exists $\theta_0 \in (0, \Theta)$ such that
\begin{equation}
\sup_{\bar D} \{ w (y; \theta_0) - \psi (x, y) \} = 0 . \label{sup*}
\end{equation}
In view of inequalities (III), $w (y; \theta_0) - \psi (x, y)$ does not vanish for
$(x, y) \in \partial D$. Moreover, as in the proof of Theorem~3, condition
\eqref{bound} that this function is separated from zero when $(x,y) \in (\RR \times
[0, \delta]) \cup \{ x \in \RR , y \in [\eta (x) - \delta, \eta (x)] \}$ for some
$\delta > 0$. According to assumption \eqref{sup*}, either there exists $(x_0, y_0)$
belonging to $D$, lying outside the $\delta$-strips described above and such that $w
(y_0; \theta_0) - \psi (x_0, y_0) = 0$ or there exists a sequence $\{(x_k,
y_k)\}_{k=1}^\infty$ located in the same strip as $(x_0, y_0)$ and such that
\begin{equation*}
v (y_k; \lambda_0) - \psi (x_k, y_k) \to 0 \quad \mbox{as} \ k \to \infty .
\end{equation*}
In both cases, we arrive to a contradiction in the same way as in the proof of
Theorem~3. Therefore, $w (y; 0) - \psi (x, y)$ is non-positive on $\bar D$ and
vanishes when $y = 0$.

We recall that $w (y; 0) = u_+ (y; s^*)$ by property (II). Since this function
satisfies property (I), we obtain that 
\[ \nabla^2 [u_+ (\cdot; s^*) - \psi] + [u_+ (\cdot; s^*) - \psi] \int_0^1 \omega'
(t [u_+ (\cdot; s^*) - \psi]) \, \D t = 0 \ \ \mbox{in} \ D .
\]
Then the maximum principle implies that $u_+ (\cdot; s^*) - \psi$ does not vanish at
inner points of this strip, that is, $u_+ (y; s^*) - \psi (x, y) > 0$ there, which
completes the proof.

\section{Discussion}

In the framework of the standard formulation, we have considered the problem
describing steady, rotational, water waves in the case when a counter-current might
be present in the corresponding flow of finite depth. Bounds on characteristics of
wave flows are obtained in terms of the same characteristics but corresponding to
some particular horizontal, shear flows that have the same vorticity distribution
$\omega$. It is important that our method allowed us to obtain bounds for stream
functions that change sign within the flow domain for which either $\check \eta$ or
$\hat \eta$ is greater than $h_0$.

It should be also mentioned that according to assertion (4) of Theorem~1 in
\cite{KKL} no unidirectional solutions exist for $r > r_0$ in the case of $\omega$
satisfying conditions (iii). Theorems~3 and 4 complement this result as follows. If
$\omega$ satisfies conditions (ii), then a wave flow such that $\check{\eta} \geq
h_0$ and $\psi \leq 1$ has a near-bottom counter-current, whereas if $\omega$
satisfies conditions (iii), then a wave flow such that $\check{\eta} > h_0$ and
$\psi \geq 0$ has a counter-current near the free surface. Thus, no unidirectional
flows exist in these cases.

An essential feature of the obtained bounds is that they, unlike those in
\cite{KN}, vary depending on the vorticity type. Indeed, inequalities (5.2a) in
\cite{KN} exclude the vorticity distributions satisfying conditions (ii) and (iii).
Another important point, that distinguishes our results from those in \cite{KN} and
also in \cite{KK4}, is that no extra requirement is imposed on $\omega$ and the
latter is assumed to be merely a locally Lipschitz function. Indeed, to prove
Theorems~3.2 and 3.4 in \cite{KK4} (they are analogous to Theorems~1 and 2 here) it
was assumed that $\mu = {\rm ess\ sup}_{\tau \in (-\infty, +\infty)}\ \omega'
(\tau)$ is less than $\pi^2 / \check \eta^2$ and $\pi^2 / \hat \eta^2$ respectively
(the last condition was also used in \cite{KN}, whereas another bound was imposed on
$\mu$ in Theorem~3.6, \cite{KK4}). However, much weaker assumption about $\omega$
satisfying conditions (ii) yields assertion (C) of Theorem~1 for a wider range of
values of the Bernoulli constant than $r \in (r_c, r_0)$, and below we outline how
to demonstrate this.

Another point concerning assertion (C) of Theorem~3.2 in \cite{KK4} should be
mentioned. The assumption that $\check \eta < h (s^>)$ (it is expressed in terms of
the notation adopted in the present paper) used in the proof of this assertion is
missing in the formulation of that theorem. A similar omission made in Theorem~3.4,
\cite{KK4}, is as follows. The assumption $\hat \eta < h (s^<)$ used in the proof
of assertion (B) of this theorem is missing in its formulation.

Let us turn to assertion (C) of Theorem~1. Since $s_0 = 0$ for $\omega$ satisfying
conditions (ii) which is assumed in what follows, the functions $U (y; s)$ and $h
(s)$ are defined by formulae \eqref{eq:Uim} and \eqref{eq:d} respectively for all $s
\geq 0$. Then the following formulae
\[ U(y; s) = U (y + 2 \, y_- (-s); -s) , \quad h (s) = h (-s) - 2 \, y_- (-s)
\]
with $y_-$ given by \eqref{eq:y_pm} extend these functions to the negative values of
$s$ belonging to some interval adjacent to zero. It is clear that both functions are
continuously differentiable, and Lemma~1 implies that $\dot h (s) < 0$ when $s < 0$
is in a neighbourhood of zero.

Let $s' \geq - s^>$ be such that $(s', 0)$ is the largest interval where $\dot h$
is negative, then
\begin{equation}
\partial_s U(y; s) > 0 \quad \mbox{for} \ y \in (0, h (s)) \ \mbox{and} \ s \in (s',
\infty) . \label{lem2}
\end{equation}
Indeed, if $U (y; s)$ is not a monotonically increasing function of $s$ for some $y
\in (0, h (s))$, then there exist small negative 
\[ s_1, s_2 > s_1 \ \mbox{and} \ y_* \in (0, h (s_2)) \ \mbox{such that} \ U (y_*;
s_1) < U (y_*; s_2) .
\]
Since $U(y; s_1) > U(y; s_2)$ for $y = h (s_2)$ and when $y > 0$ is small, there
exists $s > 0$ such that $U (y; s) \geq U(y; s_2)$ for $y > 0$ and $U(y_*; s) \geq
U(y_*; s_2)$. However, this is impossible in view of the maximum principle for
non-negative functions.

It is clear that formula \eqref{eq:calR} correctly defines $\mathcal R (s)$ for $s
\in (s', 0)$. Therefore, the stream solutions $(u_+, H_+)$ and $(u_-, H_-)$ can be
found for $r \in [r_0, r')$ in the same way as in \cite{KK3}; here $r' = \lim_{s\to
s'} {\cal R} (s)$. Now we are in a position to complement Theorem~1 by the following
assertion.

\vspace{2mm}

\noindent {\bf Proposition 1.} {\it Let $\omega$ satisfy conditions {\rm (ii)}. If
problem \eqref{eq:lapp}--\eqref{eq:bep} with $r \in (r_c, r')$ has a non-stream
solution $(\psi, \eta)$ such that $\psi \leq 1$ in $\bar D$ and $\check \eta < h (s')$,
then $\check \eta \leq H_+$ and this inequality is strict provided $\check \eta$ is
attained at some point on the free surface.}

\vspace{2mm}

\noindent {\it Sketch of the proof.} It is sufficient to prove the proposition for
$r \in (r_0, r')$, in which case there exists $\check{s} \in (s', 0)$ such that $h
(\check s) = \check{\eta}$. In the same way as in the proof of Theorem~3, one
constructs a family, say $v (y, \lambda)$, that depends on $(y, \lambda) \in (0, h
(\check s)) \times [0, \Lambda]$ continuously and satisfies properties (I)--(IV)
listed in that proof. Then applying inequality \eqref{lem2} and the definition of
$\check s$, one completes the proof using the same argument as in \S\,3.1 with $v
(y, \lambda)$ instead of $U_\ell (y)$.

\vspace{2mm}

In conclusion of this section, it remains to show that the existence of $s'$ means
that $\omega'$ satisfies the following condition. For every $s > s'$ the inequality
\begin{equation}
\int_0^{h (s)} \left[ |v' (y)|^2 - \omega'(U(y; s)) |v (y)|^2 \right] \D y > 0
\label{fin}
\end{equation}
holds for every non-zero $v$ belonging to the Sobolev space $H_0^1 (0, h (s))$.

Indeed, we have that $U (h(s); s) = 1$, and so
\[ \D U (h(s); s) / \D s = \partial_y U (h(s); s) \, \dot h (s) + \partial_s U 
(h(s); s) = 0 .
\]
Since $[\partial_y U (h(s); s)]_{s=s'} \neq 0$, the equality $\dot h (s') = 0$
implies that $z (y) = [\partial_s U (y; s)]_{s=s'}$ is a nontrivial solution of the
boundary value problem
\[ z'' + \omega' (U (\cdot; s')) \, z = 0 \ \mbox{on} \ (0, h (s')) , \quad z (0) = 
z (h (s')) = 0 .
\]
This yields the property of $\omega'$ formulated above.

If $\mu < \pi^2 / [h (s)]^2$, then inequality \eqref{fin} holds for the described
test functions, and so this property of $\omega'$ is weaker than the bound imposed
on $\mu$ in \cite{KN} and \cite{KK4}.

\vspace{6mm}

\noindent {\bf Acknowledgements.} V.~K. was supported by the Swedish Research
Council (VR). N.~K. acknowledges the support from the Link\"oping University.

\end{document}